# Integration of Artificial Intelligence in Educational Measurement: Efficacy of ChatGPT in Data Generation within the Scope of Item Response Theory

Hatice Gurdil[1], Yesim Beril Soguksu[1], Salih Salihoglu[2]* and Fatma Coskun[3]

*[1]Measurement and Evaluation, Ministry of Education, Ankara, Turkey; [2]Department of Industrial and Systems Engineering, University of Miami, Coral Gables, Florida, United States; [3]Department of Measurement and Evaluation, University of Sutcu Imam, Kahramanmaras, Turkey*

**Abstract**

The aim of this study is to investigate the effectiveness of ChatGPT 3.5 in developing algorithms for data generation within the framework of Item Response Theory (IRT) using the R programming language. In this context, validity examinations were conducted on data sets generated according to the Two-Parameter Logistic Model (2PLM) with algorithms written by ChatGPT 3.5 and researchers. These examinations considered whether the data sets met the IRT assumptions and the simulation conditions of the item parameters. As a result, it was determined that while ChatGPT 3.5 was quite successful in generating data that met the IRT assumptions, it was less effective in meeting the simulation conditions of the item parameters compared to the algorithm developed by the researchers. In this regard, ChatGPT 3.5 is recommended as a useful tool that researchers can use in developing data generation algorithms for IRT.

## 1. Introduction

ChatGPT, an artificial intelligence chatbot based on the Large Language Model (LLM), uses OpenAI's Generative Pre-trained Transformer 3.5 (GPT-3.5) language model to generate text in response to user input. The GPT-3.5 model, which is among the most sophisticated language models in the field of natural language processing (NLP), has been trained with over 175 billion parameters from various internet sources such as books, articles, web pages, and social conversations (Uc-Cetina et al., 2023). It can respond to question prompts with a trained dataset that is 570 GB in size (van Dis et al., 2023). Despite the previous use of artificial intelligence tools that transformed research applications (such as Grammarly, rTutor.ai, Research Rabbit, etc.) (Else, 2023), ChatGPT stands out from earlier models with its vast volume of raw data, multitude of parameters used in learning, context-specific architecture, and advanced features like supervised learning utilized in its development (Floridi, 2023; Susnjak, 2022). Additionally, ChatGPT, different from the previous avatars of machine learning in its ability to analyze existing data and generate new data, has become the fastest-growing consumer application in history by reaching 1 million users within the first 5 days after its launch and over 100 million users as of January 2023. It is also one of the most popular LLMs supporting code generation (Hu, 2023; Aljanabi et al., 2023; Feng et al, 2023, Khoury et al., 2023).

*Corresponding Author: Salih Salihoglu, Author Email: sxs4331@miami.edu

Despite its rapid development, ChatGPT, which constructs responses from data traces and does not rely on logic or reasoning, is probabilistic and stochastic and is limited by the quality of its training set (Bender et al., 2021; Perrigo, 2023). Moreover, since it lacks the ability to browse the internet, when the data and algorithms it depends on are flawed or biased, the outputs are likely to be flawed or biased as well (Deng & Lin, 2022, McGee, 2023, OpenAI, 2023, Ortiz, 2023a, Ortiz, 2023b, Yang, 2022). Being trained only up to 2021 data, it cannot automatically incorporate real-time information and thus cannot access current knowledge. Additionally, its use by students for direct answers without in-depth reading and critical analysis of the subject matter can suppress critical thinking, problem-solving, and creativity skills necessary in academic, professional, and real-life contexts (O'Connor, 2023), and may lead to academic misconduct (Stokel-Walker, 2022). Another criticism concerns the risk of overreliance on ChatGPT, manipulation through malicious inputs, and the potential for spreading misinformation or propaganda on widely accessible platforms in terms of security (Deng & Lin, 2022).

In response to these criticisms, the New York City Department of Education has banned the use of ChatGPT (Hirsh-Pasek & Blinkoff, 2023), and the International Machine Learning Conference (2023) has limited its use only as part of experimental analysis. Despite all these restrictions, ChatGPT, one of the world's fastest-growing technologies, clearly has the potential to fundamentally change how we access information, learn, and even conduct all kinds of work, suggesting that we are on the cusp of a societal transformation. This shift, due to the possibility that it will endanger a growing number of jobs (Dwivedi et al., 2021) and changes it will cause in individuals' work systems, often encounters user resistance (Laumer et al., 2016).

Despite these negatives and resistance, considering its benefits, ease of accessibility, and widespread use, it is predicted that banning it will not be a solution (Rosenzweig-Ziff, 2023; Springer-Nature, 2023). While acknowledging the potential destructive effects of technology, this destruction is also seen as an opportunity for scientific advancements that can further advance society. Therefore, instead of trying to neutralize digital transformation with bans, it seems more logical to focus on controlling its effects in beneficial ways and integrating them. Given that academia has already started to undergo a digital transition, it appears more appropriate to focus on the positive aspects of ChatGPT technology and to steer its negative aspects towards positive outcomes.

ChatGPT has spread to a wide range of uses, and some of its features include searching for information on various topics, creating stories and reports, writing and correcting computer code, writing and summarizing articles and chapters (Baidoo-Anu & Ansah, 2023; Else, 2023; Kundalia, 2023; Rudolph et al., 2023; Zhai, 2022), conducting tests, processing and explaining data (Hassani & Silva, 2023). These features can positively impact individuals' interactions with computers (Montti, 2022) and provide students with opportunities to enhance their learning, and increase their productivity (Dwivedi et al., 2021). However, unlike humans, ChatGPT, while able to expand its intelligence without cognitive limits, may not possess as specialized intelligence as humans due to its training data being limited to narrow fields or disciplines. In this context, users should not rely on ChatGPT as the final source; instead, they should use it as a tool to discuss and broaden perspectives. This could lead to better outcomes from artificial intelligence through a new type of collaboration that offers human-machine hybrid work opportunities, where humans guide ChatGPT based on their expertise (Mollick, 2022; van Dis et al., 2023). For this, the potential of artificial intelligence in hybrid teams must first be understood. Considering that ChatGPT can produce biased outputs and cannot verify the reality of data, it should be used as support, ensuring that it does not replace the researcher's fundamental tasks such as data analysis, interpretation, and drawing conclusions (Elsevier, 2023).

Considering that ChatGPT generates scientific knowledge from digital traces, it should be noted that it cannot progress alone without human intervention, especially in complex tasks (Biswas, 2023), and that the responsibility lies solely with the user. In this context, users should understand the need to ask the right questions and evaluate the quality of the responses, realizing that as their expertise increases, so will the quality of the information obtained from ChatGPT and their ability to interpret its outputs. Additionally, critically viewing the contributions obtained, validating their accuracy through research from different sources, or making adjustments based on examinations also fall under user responsibility.

It is believed that using ChatGPT under user responsibility will assist users in quickly accessing information and performing mundane, repetitive tasks. Consequently, users can focus on higher-level skills, thereby becoming more productive, and the process itself can become more efficient. Considering that the contributions of ChatGPT in the field of education are also increasing, and that education has been redesigned in line with technology for decades (Baidoo-Anu & Ansah, 2023; Huang, 2019), it is crucial to adapt to this transformation promptly. In this context, it is necessary to demonstrate the potential and limitations of ChatGPT, which is used in many fields. One of these broad application areas is software and programming. ChatGPT assists users in understanding and solving technical problems by providing guidance in complex topics like computer programming, programming languages, algorithms, and data structures. Offering a wide range of capabilities in these areas, ChatGPT can facilitate the processes of designing, creating, developing, testing, and maintaining software, including writing, completing, correcting, predicting, and debugging code. ChatGPT, capable of maintaining logical consistency while answering questions related to programming challenges, has extraordinary features in code generation (Chen et al., 2023; Dong et al., 2023; Liu et al., 2023; OpenAI, 2022; OpenAI, 2023). Thanks to these features of ChatGPT, researchers can improve code quality; save time and effort, thus focusing on more creative work due to reduced cognitive load, and enhance productivity (Jaber et ak., 2023; Biswas, 2023; Chen et al., 2022). ChatGPT, with such positive features, is not limited to natural language and can communicate in more than ten programming and querying languages, including C++, C#, Java, Python, R, etc. (Feng et al., 2023). When examining research in this direction, studies related to software (Adamson & Bägerfeldt, 2023; Jaber et al., 2023; Biswas, 2023), research on ChatGPT's code generation and debugging task, efficiency, and accuracy (Aljanabi, et al., 2023; Bang et al., 2023; Feng et al., 2023; Hansson & Ellréus, 2023; Kashefi & Mukerji, 2023; Sakib et al., 2023; Sobania et al., 2023; Surameery & Shakor; 2023, Tian et al., 2023; White et al., 2023) are available; however, no study examining its effectiveness in data generation has been found.

The rapid advancement of technology is also increasing the number of scientific studies and introducing new methods in various fields. In examining the effectiveness of these new methods, simulation studies are often conducted, which frequently involve comparisons with previous methods. At the same time, experimental studies are needed to demonstrate the functionality of these methods under different conditions, where some conditions are kept constant while others are varied. To conduct these experimental studies, it is first necessary to generate data under the mentioned conditions. In educational sciences, data generation is often carried out in accordance with Item Response Theory (IRT). IRT focuses on the probability of an individual's performance on an item based on their ability. IRT models, which can be falsified through model-data fit, have strong graphical and mathematical aspects (DeMars, 2010). In recent years, the R programming language, which is among the popular languages, is used in data generation related to IRT. It is known that the R language is supported by ChatGPT 3.5 (Feng et al., 2023). In this context, this study aims to determine how effective ChatGPT 3.5 is in writing code for IRT-based data generation using the R language. This work is expected to make many important contributions to the field of educational measurement and the application of artificial intelligence in data generation, and to demonstrate the potential of using ChatGPT 3.5, a state-of-the-art language model, to develop data generation algorithms for Item Response Theory (IRT). With this innovative approach, it is expected to provide a new perspective on how artificial intelligence can assist researchers in educational and psychological measurements, as well as its ability to simulate educational data. In the context of a rapidly evolving technological environment, this study underlines the importance of integrating artificial intelligence tools such as ChatGPT 3.5 into research methodologies. To evaluate ChatGPT 3.5's effectiveness in this regard, the data sets produced are analyzed to see to what extent they meet the conditions specified in the created simulation design and whether the data sets can indeed be produced as intended. Additionally, the data sets produced with the help of ChatGPT 3.5 algorithms are compared with data sets generated by algorithms developed by researchers. In this context, in this study, the validity of the data sets produced by ChatGPT 3.5 were comprehensively addressed by evaluating them according to basic IRT assumptions such as one-dimensionality, local independence and model-data fit. In this direction, the following research questions have been explored:

For data sets produced according to the Two Parameter Logistic Item Response Theory Model (2PLM) using ChatGPT 3.5 algorithms;

- Do they meet the unidimensionality assumption?
- Do they meet the local independence assumption?
- Do they meet the model-data fit assumption with the 2PLM?
- How many item parameters fall outside the ranges specified in the simulation design?
- What are the bias and RMSE values of the item parameters?

## 2. Methodology

### 2.1. Research model

In the study, three different algorithms are used for data generation, and it is investigated to what extent the data sets developed with these algorithms meet the conditions specified in the simulation pattern. Two of the algorithms developed in the R language are produced by ChatGPT 3.5 in October 2023, while the third algorithm is developed by the researchers. Using these developed algorithms, data sets are generated, followed by validity examinations of the generated data sets. In this aspect, the study is experimental in nature (Morris et al., 2017).

### 2.2. Simulation pattern

In this study, the Monte Carlo simulation used in previous simulation studies (Cohen et al., 1996; Li et al., 2012) is employed to investigate the effects of various altered conditions on ChatGPT 3.5's performance in data generation validity. The Monte Carlo approach allows for the distribution of parameter estimates to be obtained, and reduces the chance of deriving unreasonable results from a single data set. In this context, multiple repetitions (replications) are conducted to resample the true parameters (Harwell et al., 1996). In thi study, the recommended number of repetitions for accurate and reliable parameter estimations (Feinberg & Rubright, 2016) is considered, and 100 repetitions are conducted for each data set generated according to the specified simulation conditions. While determining the fixed and altered conditions of the simulation pattern, studies from the literature are referred to. The 4.3.2 version of the R programming language is used in conducting the study. All the experimental conditions used in the study are presented in Table 1.

**Table 1.** Simulation Pattern

|  | Conditions | Levels | Number of Levels |
|---|---|---|---|
| Fixed Conditions | IRT Model | Unidimensional 2PL | 1 |
|  | Ability ($\theta$) | $-3 \leq \theta \leq 3$ | 1 |
|  | Item Discrimination (a) | $1 \leq a \leq 2$ | 1 |
|  | Item Difficulty (b) | $-2 \leq b \leq 2$ | 1 |
| Altered Conditions | Item Number (k) | 20, 40 | 2 |
|  | Sample Size (n) | 500, 2000 | 2 |
| Total Number of Conditions |  |  | 2x2=4 |
| Number of Repetitions |  |  | 100 |
| Total Data Sets |  |  | 4x100=400 |

When examining Table 1, it is observed that different conditions have been considered for the production of unidimensional data for items scored dichotomously (0-1) within the scope of the study. The IRT model uses the 2PLM (DeMars, 2010) for dichotomously scored items, the individual ability parameter is frequently drawn from a normal distribution with a mean of 0 and standard deviation of 1 (Feinberg & Rubright, 2016), item discrimination parameters are taken from a uniform distribution with a minimum of 1 and maximum of 2, which are thought to be more suitable for 2PLM estimates and fall within the ranges defined in the literature (Dekker, 2004, Hambleton et al., 1991), and item difficulty is also set from a uniform distribution with a minimum of -2 and a maximum of 2, in accordance with the ranges determined in the literature (DeMars, 2010). In Monte Carlo studies, considering that more accurate item parameter estimates can be made in tests with an item number greater than 20 and sample size greater than 500 (De Ayala, 2009; Stone, 1992), the number of items has been altered to 20 (short) and 40 (long), and the sample size has been varied to 500 (small) and 2000 (large).

The first and second algorithms (A1 and A2) are developed by ChatGPT 3.5 through prompts for data generation. Different prompts are used to request ChatGPT 3.5 to produce new algorithms; however, when it is determined that the generated algorithms do not differ from A1 and A2, the algorithm development process with ChatGPT 3.5 is terminated. The prompts used for A1 and A2 are given in Appendix 1 and 2. The third algorithm (A3) used for data generation is developed by researchers using the *'mirt'* package (Chalmers, 2012) in the R language. The codes for the three developed algorithms are located in Appendices 3, 4, and 5. During the data generation phase, the developed algorithms are run in the R environment, and data sets with 100 repetitions are produced considering the conditions in the simulation pattern. As a result, a total of 2x2x100=400 different data sets are produced, comprising 2 different test lengths (20, 40), 2 different sample sizes (500, 2000), and 100 repetitions.

**2.3. Data analysis**

Within the scope of the research, data sets generated sequentially with A1, A2, and A3 are expected to meet the assumptions (unidimensionality, local independence, model-data fit) specific to the unidimensional 2PLM, as they are generated according to this model. Otherwise, the generated data will not conform to the intended data production scenario, and consequently, the results obtained will be erroneous. Furthermore, the parameters estimated in the generated data sets must meet the conditions specified in the simulation pattern to demonstrate the accuracy of the data generations. Therefore, the first step is to check whether the produced data sets comply with the IRT assumptions and whether the parameters meet the conditions specified in the simulation pattern, and for this purpose, the following examinations are conducted.

**Unidimensionality Examination of Data Sets:** To examine the unidimensionality assumption, Exploratory Factor Analysis (EFA) is conducted using the *'psych'* (Revelle, 2020) and *'sirt'* (Robitzsch, 2019) packages available in the R programming language. In this context, *'tetrachoric correlation matrices'* suitable for dichotomously scored (1-0) data sets are created for all data sets. Within the scope of EFA assumptions, the results of the Kaiser-Meyer-Olkin (KMO) and Bartlett tests are examined. For both A3 and the data sets produced with the help of ChatGPT 3.5 (A1 and A2) algorithms, it is found that the KMO test results are greater than 0.87 and the Bartlett tests are significant (p<.05) for each condition. After meeting the assumption of unidimensionality, the number of factors is determined using Horn 's *'Parallel Analysis'* method (Horn, 1965). The *'fa.parallel'* function from the *'psych'* package is utilized for the factor extraction process; eigenvalues for the factors and scree plot graphs are examined. In deciding on the number of factors, factor loadings, explained variance ratios, and scree plots are taken into account.

**Local Independence Examination of Data Sets:** The Item Response Theory (IRT) local independence assumption, which states that item responses should be independent of each other when individuals' abilities are controlled for (DeMars, 2010), has been checked using Yen's Q3 test (Yen, 1984). To verify the local independence assumption, the *'Yen.Q3'* function from the *'subscore'* package (Dai et al., 2022) in the R language is utilized. The condition set by (Yen, 1984) for local independence issues, where correlations between residuals for problematic items exceeding 0.20, is taken as the criterion for identifying violations.

**Model-Data Fit Examination of Data Sets:** In checking the compatibility of the generated data sets with the 2PLM, the significance levels of the M2 statistic are examined. If the M2 statistic, constructed using the residuals between observed and expected marginal probabilities, is not significant, it is assumed that the data set is consistent with the model (Maydeu-Olivares & Joe, 2006). The *'M2'* function from the *'mirt'* package (Chalmers, 2012) in the R programming language is used to check the significance of the M2 statistic. Additionally, in assessing model-data fit, the nested IRT models' (1PL, 2PL, 3PL) *'-2 log-likelihood'* values and significances was taken into account (Thissen & Steinberg, 1986). The *'anova'* function in the R programming language is utilized for comparing the models.

**How Many Item Parameters Fall Within the Specified Ranges:** It is determined how many of the item discrimination (a) and item difficulty (b) parameters in the generated data sets meet the conditions specified in the simulation pattern. In this context, the researchers create counters in the R programming language to identify the number of parameters that fall outside the specified ranges. Considering the altered item numbers and number of repetitions in the research, it is established how many item difficulty and item discrimination parameters fall outside the specified ranges, with 20 items for 20x100=2000 and 40 items for 40x100=4000.

**Bias and Error (RMSE) Values of Item Parameters:** In assessing the bias and error (RMSE) values of the item parameters for data sets generated considering the conditions specified in the simulation pattern, the formulas located in Equations 1 and 2 are used.

$$Bias = \frac{\sum_{i=1}^{K}(\hat{X}_i - X_i)}{K} \qquad (1)$$

$$RMSE = \sqrt{\frac{\sum_{i=1}^{K}(\hat{X}_i - X_i)^2}{K}} \qquad (2)$$

In Equations 1 and 2, $\hat{X}_i$ represents the parameter for item i $(i = 1, 2, ..., K)$, $X_i$ denotes the actual parameter estimate, and $K$ indicates the number of items. Bias and RMSE values are calculated for each condition in the data sets with 100 repetitions, and then their averages are taken.

## 3. Findings

This section presents the findings obtained within the scope of the research. In the study, data generation for the 2PLM is carried out using three different algorithms (A1, A2, and A3) based on the conditions specified in the simulation pattern. Examinations are conducted to determine the extent to which the generated data sets meet the specified simulation conditions. The results of these examinations are shown in Table 2.

**Table 2.** Results of the Analyses on the Data Sets

| Condition | | Number of Items | Sample Size | Number of Factors | Parallel Analysis (Number of Factors) | | Local Independence Violation Yen Q3 | M2 Fit Violation | Nested Model Fit | Out of Range Parameter a | | Out of Range Parameter b | | Average Bais a | Average Bais b | Average RMSE a | Average RMSE b |
|---|---|---|---|---|---|---|---|---|---|---|---|---|---|---|---|---|---|
| 1 | A1 | 20 | 500 | 1 | 1 | (4 x 2) | 13 | 0 (%0)* | 2PL | 1549 | (%77)** | 0 | (%0)*** | -1.201 | -0.093 | 2.108 | 1.145 |
| 2 | A2 | 20 | 500 | 1 | 1 | | - | 5 (%5) | 2PL | 241 | (%12) | 632 | (%32) | 0.012 | 0.206 | 0.21 | 1.936 |
| 3 | A3 | 20 | 500 | 1 | 1 | | - | 4 (%4) | 2PL | 222 | (%11) | 86 | (%4) | 0.018 | 0.04 | 0.189 | 0.137 |
| 4 | A1 | 20 | 2000 | 1 | 1 | (4 x 2) | 2 | 4 (%4) | 2PL | 1580 | (%79) | 0 | (%0) | -1.227 | 0.397 | 2.091 | 1.037 |
| 5 | A2 | 20 | 2000 | 1 | 1 | | - | 4 (%4) | 2PL | 78 | (%3) | 661 | (%33) | 0 | -0.407 | 0.103 | 2.042 |
| 6 | A3 | 20 | 2000 | 1 | 1 | | - | 4 (%4) | 2PL | 84 | (%4) | 45 | (%2) | 0.007 | -0.288 | 0.093 | 0.057 |
| 7 | A1 | 40 | 500 | 1 | 1 | | 9 | 4 (%4) | 2PL | 3129 | (%78) | 0 | (%0) | -1.297 | -0.013 | 2.142 | 1.215 |
| 8 | A2 | 40 | 500 | 1 | 1 | | - | 4 (%4) | 2PL | 495 | (%12) | 1280 | (%32) | 0.019 | -0.073 | 0.198 | 2.093 |
| 9 | A3 | 40 | 500 | 1 | 1 | | - | 7 (%7) | 2PL | 427 | (%11) | 173 | (%4) | 0.013 | 0.012 | 0.183 | 0.159 |
| 10 | A1 | 40 | 2000 | 1 | 1 | | - | 6 (%6) | 2PL | 3181 | (%79) | 0 | (%0) | -1.271 | 0.421 | 2.16 | 1.155 |
| 11 | A2 | 40 | 2000 | 1 | 1 | | - | 2 (%2) | 2PL | 135 | (%3) | 1248 | (%31) | 0.003 | -0.186 | 0.095 | 2.133 |
| 12 | A3 | 40 | 2000 | 1 | 1 | | - | 7 (%7) | 2PL | 144 | (%4) | 79 | (%2) | 0.002 | 0.086 | 0.09 | 0.077 |

*Percentage of fit violation  ** Percentage of a parameter outside the range  *** Percentage of a parameter outside the range

As can be seen in Table 2, it is found that in all conditions where the number of items is 20, the EFA analysis indicates the presence of a dominant single dimension in the data sets with 100 repetitions. In the parallel analysis, only in A1, two conditions (20 items with 500 samples, 20 items with 2000 samples) reveal a two-factor structure for 4 data sets. In all conditions where the number of items is 40, both EFA and Parallel Analysis results support a single-factor structure. Consequently, it can be said that the data sets generated with ChatGPT 3.5 algorithms are unidimensional as expected, with a few negligible exceptions.

Upon examining Table 2, it is observed that violations of the local independence assumption are seen only in some cases with the ChatGPT 3.5 A1 algorithm. In this algorithm, violations of local independence in item pairs are identified in 13 out of 100 data sets for 20 items with 500 samples, in 2 data sets for 20 items with 2000 samples, and in 9 data sets for 40 items with 500 samples. At this point, considering the absence of data sets with local independence violations, it can be said that the ChatGPT 3.5 A2 algorithm and A3 are more successful.

It is observed that, according to the significance checks of the M2 statistic, the proportion of data sets that do not meet the model-data fit varies between 0% and 7% for all algorithms upon examining Table 2. In conditions with 20 items and a sample size of 500, the best model-data fit is achieved with A1, with no fit violation observed (0%). When the total number of items is 40 and the sample size is 2000, it is determined that the least model-data fit violation (2%) is with A2, while the most violations (7%) occur with A3. Additionally, considering the nested model fit, it is determined that in all conditions, the data sets are better fitted with the 2PL model. Generally, it can be said that ChatGPT 3.5 algorithms (A1 and A2) are at least as successful as the researcher-developed A3 in terms of the compatibility of the data sets with the 2PL model.

Furthermore, it is evident that the data generation for the parameter a, fitting the ranges specified in the simulation pattern, is successfully conducted with ChatGPT 3.5 A2 and the researcher-developed A3 across all conditions. ChatGPT 3.5 A1, however, appears quite weak in generating data within the specified ranges for the parameter a, with a majority of the produced a parameters (77%-79%) falling outside the determined ranges. In all conditions with a sample size of 500, A3 shows the least percentage of data generation outside the specified ranges for the parameter a, while in conditions with a sample size of 2000, A2 has the least percentage of deviation. For the parameter b, the best data generation within the specified ranges is achieved with A1, followed closely by A3. A2, however, is weak in generating data within the specified ranges for parameter b, with a portion of the b parameters (31%-33%) falling outside the specified ranges. In conclusion, although successful in meeting the IRT assumptions, it can be said that the item parameter estimations in the data sets generated with ChatGPT 3.5 algorithms have more difficulties in meeting the conditions specified in the simulation pattern compared to A3.

As for bias average, it is observed that for the parameter a across all conditions, the lowest bias average is found in A1, while A2 and A3 produce similar values. In the condition with a total item count of 20 and a sample size of 500, the lowest bias average for parameter b is in A1, while in all other conditions, it is in A2. In conditions with a sample size of 2000, the highest bias average is obtained with A1. For conditions with a sample size of 500, the highest bias average for a total item count of 20 is with A2, and for 40, it is with A3. Additionally, the highest RMSE average values across all conditions are determined to be for parameter a in A1 and for parameter b in A2. Furthermore, across all conditions, the lowest RMSE average values for both parameters a and b are found in A3. Therefore, it can be said that the best results in terms of bias and RMSE averages for item parameters are obtained in data sets produced with A3, while ChatGPT 3.5 algorithms lead to higher bias and RMSE averages.

## 4. Conclusion and discussion

When the findings are evaluated, it can be stated that ChatGPT 3.5 algorithms are generally successful in generating data that conforms to IRT assumptions. Excluding a few negligible exceptions, unidimensional data sets with no local independence violations and compatible with the 2PLM are generated using ChatGPT 3.5 algorithms. However, when evaluating in terms of the compliance of item parameters with the ranges specified in the simulation pattern and the associated bias and RMSE averages, it appears that ChatGPT 3.5 algorithms have more issues compared to the algorithm developed by researchers.

Based on the results in Table 2, it is found that in all conditions, EFA analysis of the data sets generated with the three algorithms determines a unidimensional structure, as specified in the simulation pattern. In the parallel analysis, except for two conditions in A1, which account for a total of eight data sets, a unidimensional structure is observed in all conditions. The local independence checks show that the assumption is met in all conditions except for three in A1. M2 fit violations in all three algorithms reveal the presence of violating data sets, but these violations are relatively few, varying between 0% and 7%. Generally, the data sets that best fit the 2PL are those produced with A1 for 20 items with 500 samples, and those produced with A2 for 40 items with 2000 samples, with no clear pattern of fit violations among the algorithms. When nested models are compared for fit with 2PL, data sets in all algorithms are found to align well with 2PL. In the examination of item parameters a and b falling outside the specified ranges, it is seen that the closest results to the specified ranges are obtained with A3. The A1 algorithm is found to be inadequate in producing a parameters within the desired ranges, and A2 is inadequate for b parameters. The bias and RMSE averages for these parameters are found to support this result as expected.

In conclusion, data generation using ChatGPT 3.5 algorithms is generally successful, but the best compliance with the conditions of the simulation pattern is achieved with the algorithm developed by researchers (A3). Data sets generated with the support of ChatGPT 3.5 algorithms are found to be less adequate than those generated by researchers without ChatGPT 3.5 support, especially in terms of generating unbiased parameters facilitate the expert's work within the desired ranges. Similar to literature findings, the findings reveal that ChatGPT can successfully comply with IRT assumptions and underline the indispensable role of human expertise in driving AI outcomes.

Similar to literature findings, ChatGPT 3.5 was found to be not perfect at generating computer code and ChatGPT 3.5's solutions were found to be of lower quality than human solutions (Kundalia, 2023; Adamson & Bägerfeldt, 2023). However, there are also studies stating that ChatGPT 3.5 can produce correct solutions to many problems but falls short in some aspects, similar to this study (Tian et al., 2023; Hansson & Ellréus, 2023; Sakib et al., 2023). Additionally, it is found that there are problems in meeting unidimensionality and local independence assumptions in one of the data algorithms produced with the support of ChatGPT 3.5 (A1), while these problems are not observed in the other (A2). This may be attributed to the fact that the two algorithms are developed by different researchers using different prompts. As a result of this study, it can be said that ChatGPT 3.5 can produce data close to an expert level, as stated in some studies in the literature (Surameery & Shakor, 2023) and the accuracy of the generated data set will increase in line with the expertise. As mentioned in the studies (Bang et al., 2023), the success of algorithms developed with ChatGPT 3.5 is influenced by the prompts and guidance provided by experts. It can be stated that as expertise increases, the quality of prompts entered into ChatGPT 3.5 also increases, thus bringing the output closer to expert accuracy. As seen, contrary to fears, ChatGPT 3.5 does not eliminate the need for expertise; rather, it serves as a tool to facilitate the expert's work. In addition, it should be taken into consideration that ChatGPT has certain capabilities and limitations in each version. In this study, the adequacy of ChatGPT 3.5, which is free of charge and accessible to more people, in developing algorithms for data generation within the scope of IRT was examined. In the validity checks carried out within the scope of the study, it was observed that in the data sets produced with ChatGPT 3.5 algorithms, the item parameters were weak in meeting the ranges specified in the simulation pattern and local independence violations were more common. ChatGPT 4 version or in the future versions, much better results can be obtained than those obtained in the research. Because it is possible that the nature of the version used may have a direct impact on the reliability and accuracy of the results obtained in the research. It should be taken into account that more recent versions may tend to produce more accurate results as they are trained on a larger training data set.

In a world where technology is increasingly advancing and playing a major role, banning technology is clearly not meaningful (King, 2023). In line with all these results, it seems more appropriate to use ChatGPT 3.5 as a support tool in research, as suggested in the studies (Hansson & Ellréus, 2023; Sollie, 2009, Biswas, 2023; Deng & Lin, 2022, Dwivedi et al., 2023, Surameery & Shakor, 2023), rather than restricting its use. The results of this study reveal a balanced approach in which artificial intelligence is used as a supporting tool to enhance human capabilities rather than replace them. It is envisioned that this integration could lead to more efficient research processes, allowing researchers to focus on higher-level analytical tasks while relying on AI for data generation and preliminary analysis. In this direction, considering the importance of expertise in the use of ChatGPT 3.5, it is advisable for researchers to benefit from ChatGPT 3.5 in data generation and shorten the time they spend on data generation. Researchers can use the remaining time for more creative parts of their research, as suggested in the studies (Hirsh-Pasek & Blinkoff, 2023). The situation of different outputs being produced depending on the commands of different experts can be examined in a different study that takes into account inter-expert variations. Additionally, it is recommended to conduct further in-depth studies to examine the fact that ChatGPT 3.5 generally meets IRT assumptions but does not fully achieve item recovery.

**Declarations**

Conflict of interest: The authors have no relevant financial or non-financial interests to disclose. The authors have no conflicts of interest to declare that are relevant to the content of this article.

Data availability: Data is not publicly available, but anonymised data sets are available on request.

Stone, C. A. (1992). Recovery of marginal maximum likelihood estimates in the two-parameter logistic response model: An evaluation of MULTILOG. *Applied Psychological Measurement, 16(*1), 1-16.

Surameery, N. M. S., & Shakor, M. Y. (2023). Use chat gpt to solve programming bugs. *International Journal of Information Technology & Computer Engineering (IJITC) ISSN: 2455-5290*, *3*(01), 17-22. https://doi.org/10.55529/ijitc.31.17.22

Susnjak, T. (2022). ChatGPT: The End of Online Exam Integrity? *arXiv preprint arXiv: 2212.09292*. https://doi.org/10.48550/arXiv.2212.09292

Thissen, D., & Steinberg, L. (1986). A taxonomy of item response models. *Psychometrika*, *51*(4), 567-577.

Tian, H., Lu, W., Li, T. O., Tang, X., Cheung, S. C., Klein, J., & Bissyandé, T. F. (2023). Is ChatGPT the Ultimate Programming Assistant--How far is it?. *arXiv preprint arXiv:2304.11938*. https://doi.org/10.48550/arXiv.2304.11938

Uc-Cetina, V., Navarro-Guerrero, N., Martin-Gonzalez, A., Weber, C., & Wermter, S. (2023). Survey on reinforcement learning for language processing. *Artificial Intelligence Review, 56*(2), 1543-1575.

van Dis, E. A. M., Bollen, J., Zuidema, W., van Rooij, R., & Bockting, C. (2023). ChatGPT: five priorities for research. *Nature, 614*, 224–226. https://doi.org/10.1038/d41586- 023-00288-7

White, J., Hays, S., Fu, Q., Spencer-Smith, J., & Schmidt, D. C. (2023). Chatgpt prompt patterns for improving code quality, refactoring, requirements elicitation, and software design. *arXiv preprint arXiv:2303.07839*. https://doi.org/10.48550/arXiv.2303.07839

Yang, S. (2022). The Abilities and Limitations of ChatGPT. Anaconda Perspectives. 〈https ://www.anaconda.com/blog/the-abilities-and-limitations-of-chatgpt〉

Yen, W. M. (1984). Effects of local item dependence on the fit and equating performance of the three-parameter logistic model. *Applied Psychological Measurement, 8*(2), 125-145.

Zhai, X., (2022). ChatGPT User experience: Implications for education. (December 27, 2022). Available at SSRN: https://ssrn.com/abstract=4312418 or http://dx.doi.org/10.2139/ssrn.4312418.
25

## 6. Appendix

*Appendix 1. Prompt 1*

1- I want to generate 1-0 response data for 2 Parametric Logistic Item Response Model. My test length is 20. My sample size is 500. My ability parameters have a normal distrubition with mean 0, standard deviation 1. Item discrimination parameter will have a uniform distrubition and between 1 and 2. My difficulty parameter (b) will have a uniform distrubition and between -2 and 2. My replication number is 100. I want to save 1-0 responses to an empty list. Can you give me the codes of this simulation in R language?

3- Ok, but this code does not label the columns

4- Ok. I want to estimate 2PL model parameters. First I want to estimate ability parameters via EAP method. Then I want to estimate item parameters a and b. Can you give me the codes?

5- R gives error about eap estimation code

6- Ok. One more thing. I have to save all of the simulated ability, difficulty and discriminatiom parameters. So the code should give me the opportunity of this.

7- No, not this. You know we simulated 1-0 response data. We generate ability, difficulty and discrimination parameters for this purpose. I need these parameters. Because after estimating model parameters I'll estimate bias and RMSE.

8- can you label the items like 'Item1', 'Item2'.

\* **Created on October 23, 2023 in ChatGPT 3.5**

*Appendix 2. Prompt 2*

1- Define the parameters that test length is 40, sample size is 2000, mean ability is 0, Sd ability is 1, minimum discrimination is 1 and maximum discrimination is 2, minimum difficulty is -2 and maximum difficulty is 2 and replications is 100 for generate 1-0 item response data for 2 Parametric Logistic Item Response Model. Generate ability parameters, discrimination parameters and item difficulty parameters. Can you give me this simulation codes in R language.

2- This code will generate ability parameters, discrimination parameters, and item difficulty parameters for each replication, calculate the probability of a correct response using the 2PL model for each item, simulate binary responses (0 or 1) based on the probabilities, assign column names to the response matrix, and store the response as a data frame in the list.

3- create a list named 'item discrimination values' to store discrimination values for each iteration in this list. And create a list named 'item difficulty values' to store discrimination values for each iteration in this list. And also create a list named 'ability values (theta)' to store discrimination values for each iteration in this list

4- I want code for added the creation of a list named simulated_parameters_list to store the simulated parameters (ability, discrimination, and difficulty) for each iteration of the loop.

5- I want code which calculates the probability of a correct response using the 2PL model and Simulate binary responses (0 or 1) based on the probability

6- I want from you to give column names to the response matrix store the response matrix directly in the list item_responses for each iteration.

**\* Created on October 23, 2023 in ChatGPT 3.5**

*Appendix 3. Algorithm 1 (A1)*

```r
library(mirt)
library(eRm)

# Parameters
test_length <- 20
sample_size <- 1000
replications <- 100
mean_ability <- 0
sd_ability <- 1
min_discrimination <- 1
max_discrimination <- 2
min_difficulty <- -2
max_difficulty <- 2
response_list=list()
# Create an empty list to store the simulated parameters for each replication
simulated_parameters_list <- list()

# Loop through each replication
for (i in 1:replications) {
  # Simulate ability parameters from a normal distribution
  ability <- rnorm(sample_size, mean_ability, sd_ability)
  
  # Simulate discrimination parameters from a uniform distribution
  discrimination <- runif(test_length, min_discrimination, max_discrimination)
  
  # Simulate difficulty parameters from a uniform distribution
  difficulty <- runif(test_length, min_difficulty, max_difficulty)
  
  # Store the simulated parameters in the list
  simulated_parameters_list[[i]] <- list(
    ability = ability,
    discrimination = discrimination,
    difficulty = difficulty
  )
  # Simulate responses using the 2PL model
  prob <- plogis(outer(ability, -difficulty, "*") * discrimination)
  responses <- matrix(rbinom(sample_size * test_length, 1, prob), nrow = sample_size)
  # Extract the response data for the current replication
  response_df <- as.data.frame(responses)
  colnames(response_df) <- paste("Item", 1:test_length)
  
  # Store the simulated response data in the list
  response_list[[i]] <- response_df
}
```

*Appendix 4. Algorithm 2 (A2)*

```r
# Install and load necessary packages
install.packages("rlist")
library(rlist)
library(mirt)

set.seed(123)

# Define the parameters
test_length <- 40
sample_size <- 2000
mean_ability <- 0
sd_ability <- 1
min_discrimination <- 1
max_discrimination <- 2
min_difficulty <- -2
max_difficulty <- 2
replications <- 100

simulated_parameters_list <- list()

# Initialize empty lists to store parameters and ability values
item_discrimination_values <- list()
item_difficulty_values <- list()
theta_values <- list()
item_responses <- list()

# Generate data for each combination
for (i in 1:replications) {
  # Generate ability parameters
  ability <- rnorm(sample_size, mean = mean_ability, sd = sd_ability)

  # Generate discrimination parameters
  discrimination <- runif(test_length, min = min_discrimination, max =
      max_discrimination)

  # Generate item difficulty parameters
  difficulty <- runif(test_length, min = min_difficulty, max = max_difficulty)

  item_discrimination_values[[i]] <- discrimination
  item_difficulty_values[[i]] <- difficulty

  # Calculate ability values (theta) for each item
  theta_values[[i]] <- ability

  # Store the simulated parameters in the list
  simulated_parameters_list[[i]] <- list(
    ability = ability,
    discrimination = discrimination,
```

```r
    difficulty = difficulty
  )

  # Initialize a matrix to store responses for this iteration
  responses <- matrix(0, nrow = sample_size, ncol = test_length)

  for (j in 1:test_length) {
    # Calculate the probability of a correct response using the 2PL model
    p_correct <- 1 / (1 + exp(-discrimination[j] * (ability - difficulty[j])))

    # Simulate binary responses (0 or 1) based on the probabilities
    responses[, j] <- rbinom(sample_size, 1, p_correct)
  }

  # Add column names to the response matrix
  colnames(responses) <- paste("Item", 1:test_length)

  # Store the response matrix for this iteration
  item_responses[[i]] <- as.data.frame(responses)
}
```

## Appendix 5. Algorithm 3 (A3)

```r
library (mirt)
# Create empty lists to store the simulated parameters and datasets for each replication
a=list()
b=list()
ability=list()
datasets=list()
# Simulate parameters and use them to generate responses for the 2PL model
for (i in 1:100) {
    a[[i]] <-as.matrix(round(runif(20, min=1, max=2),2), ncol=1)
    b[[i]] <-as.matrix(round(runif(20, min=-2, max=2),2), ncol=1)
    ability[[i]] <-as.matrix(round(rnorm(1000, mean=0, sd=1),2), ncol=1)
    datasets[[i]] <-simdata(a=a[[i]], d=b[[i]], N=1000, itemtype='dich', Theta=ability[[i]])
    datasets[[i]] <-as.data.frame(datasets[[i]])
}
```